# Fast spin precession and strong perpendicular magnetic anisotropy in ferrimagnetic Mn4N thin films improved by Pd buffer layer

Yao Zhang[1,2], Yun Kim[3], Peter P. Murmu[4], Dingbin Huang[3], Deyuan Lyu[5],

Jian-Ping Wang[5], Xiaojia Wang[3], Simon Granville[1,2]

[1]Robinson Research Institute, Victoria University of Wellington, P.O. Box 33436, Petone 5046, New Zealand

[2]MacDiarmid Institute for Advanced Materials and Nanotechnology, P.O. Box 600, Wellington 6140, New Zealand

[3]Department of Mechanical Engineering, University of Minnesota, Minneapolis, MN 55455, USA

[4]National Isotope Centre, GNS Science, PO Box 30368, Lower Hutt 5010, New Zealand

[5]Department of Electrical and Computer Engineering, University of Minnesota, Minneapolis, MN 55455, USA

Ferrimagnets take the advantages of both ferromagnets and antiferromagnets making them promise for spintronic applications. Here we prepared ferrimagnetic Mn4N thin films with high Curie temperature and investigated the crystalline structure and magnetic properties affected by the Pd buffer layer. We demonstrated that both crystalline quality and perpendicular magnetic anisotropy (PMA) of Mn4N thin films are enhanced significantly due to the relaxation of tensile stress induced by the Pd buffer layer. We also demonstrated a fast spin precession at room temperature, almost 100 GHz, in Mn4N thin films. With the characteristics of high thermal stability, enhanced PMA by buffer layer and fast spin precession, Mn4N thin film is a promising material for spintronic applications.

## Introduction

Ferrimagnet is a type of magnetic material in which the spin sublattices antiparallel with each other but with net magnetization [Proc. Phys. Soc. A 65, 869 (1952).]. In ferrimagnets, the magnetization can be easily controlled and detected like ferromagnet, and the spin dynamics is fast with a weak stray field like antiferromagnet [Nature Materials 21, 24–34 (2022)]. Thus, ferrimagnets take the advantages of both ferromagnets and antiferromagnets and have a great potential for high-speed and high-density spintronic devices [Appl. Phys. Rev. 10, 011301 (2023)].

In all ferrimagnets, rare earth-transition metal (RE-TM) amorphous alloys where the RE and TM sublattices are coupled antiferromagnetically, such as GdFeCo and TbCo, have been widely investigated due to the tunability of composition and magnetic properties [Appl. Phys. Lett. 116, 110501 (2020)]. For instance, their exhibit perpendicular magnetic anisotropy (PMA) which makes them good for magneto-optical recording and spintronics memory devices [J. Phys. D 34, R87–R102 (2001), Appl. Phys. Rev. 10, 011301 (2023)]. Also, by tuning the composition of RE and TM, one can obtain the ferrimagnet with magnetization and angular momentum compensation point where magnetization and angular momentum equals to zero, respectively [Nature Materials 21, 24–34 (2022)]. As the spin dynamic is fast at the angular momentum compensation point, extremely fast domain wall motion driven by magnetic field and electrical current have been realized [Nature Materials 16, 1187–1192 (2017), Nature Nanotechnology 13, 1154–1160 (2018)]. However, the magnetization of RE-TM is very sensitive to the temperature [Adv. Electron. Mater. 2022, 8, 2100772]. Above 400 K, they can be crystalized and lose the amorphous structure, which ruins the magnetization. Thus, exploring ferrimagnets with high thermal stability is required for practical applications.

Mn4N is a type of RE-free ferrimagnet which exhibits a very high Curie temperature ($T_c$ = 740 K) [Phys. Rev. 125, 1893]. It shows anti-perovskite cubic structure where the Mn atoms $Mn_I$ located in the corner and $Mn_{II}$ located at the face center are antiferromagnetically coupled. On the (111) plane, there is a triangular noncollinear antiferromagnetic coupling arising from the $Mn_{II}$ atoms [1979 J. Phys. F: Met. Phys. 9 2431, Acta Materialia 234 (2022) 118021]. Furthermore, PMA can be observed in Mn4N thin films because the in-plane tensile strain from the substrate breaks the cubic symmetry [*J. Appl. Phys.* 115, 17A935 (2014), PHYSICAL REVIEW B 106, L060409 (2022)]. Recently, by doping transition metals, such as Ni, the magnetization and angular momentum compensation point can be achieved, and thus fast domain wall motion driven by electrical current has been demonstrated [Nano Lett. 2021, 21, 2580−2587]. All these excellent properties make it potentially good for spintronic applications [DOI 10.1088/1361-6463/ac292a]. Usually, Mn4N (lattice constant a=0.386nm) thin films are epitaxially grown on MgO substrate (a=0.421 nm). But this 8.3% lattice mismatch can induce a tetragonal structure with misfit dislocations, and thus affects the magnetic properties. [Appl. Phys. Lett. 105, 072410 (2014)].

On the other hand, the spin dynamics behavior of magnetic thin films is crucial for the speed of spintronic devices [Rev. Mod. Phys. 82, 2731]. Theoretically, the spin precession frequency in ferrimagnet can reach tens of GHz [Low Temp. Phys. 45, 935–963 (2019)], which is higher than that of ferromagnet [Appl. Phys. Lett. 111, 032401 (2017)]. Especially, at angular momentum compensation point, the frequency of ferrimagnet can increase significantly due to the antiferromagnetic coupling [PHYSICAL REVIEW B 73, 220402(R) (2006)]. Thus, it is interesting to explore the spin dynamics of Mn4N thin film. Yet, there is little known research about this behavior.

Here, we employed Pd as the buffer layer to relax the lattice mismatch from MgO substrate. An improvement of the crystalline quality of Mn4N thin films was observed, which turns out enhancing the PMA. Furthermore, we demonstrated a fast spin precession, ~100 GHz, in Mn4N thin films.

**Experimental methods**

Samples preparation

Mn4N thin films were epitaxially deposited on MgO(001) substrates in a Kurt J Lesker CMS-18 magnetron sputtering system with a base pressure below $3 \times 10^{-8}$ Torr. Before fabricating thin films, substrates were cleaned with an Ar plasma, and then annealed at 400 °C for 30 mins in the vacuum chamber. Both Pd and Mn4N were grown at 400 °C with DC magnetron sputtered while rotating the sample holder. Pd was grown at 20 W, 2 mTorr under Ar gas with a growth rate of 0.4 Å s$^{-1}$. Mn4N is sputtered from a Mn target at 80 W, 7 mTorr under a N2/Ar ratio of 7/43 with a growth rate of 0.5 Å s$^{-1}$. All samples were post-annealed in situ at 400 °C for 5 mins to improve the lattice structure. After cooling down to ambient temperature, a 3-nm protective Ta layer was deposited on the top.

Thin-film characterization

The crystalline structure of Mn4N thin films was characterized by X-ray diffraction (XRD) with Co Kα radiation (λ = 1.7889 Å) using Rigaku Smartlab.

Ion beam analysis (IBA) techniques such as Rutherford backscattering spectrometry (RBS) and nuclear reaction analysis (NRA) were employed to study the thickness and composition of $Mn_4N$ film [ref]. RBS measurements conducted by employing 1.0 mm collimated 2.0 MeV $^4He^+$ beam with backscattered particles collected using a surface barrier detector placed at 165° [ref]. RBS data were fitted using data deconvolution software Rutherford universal manipulation programme (RUMP) to extract film thickness and composition. NRA measurements were carried out using 920 keV $^2H^+$ beam with elastically scattered deuteron collected using the surface barrier detector equipped with 10 μm mylar filter placed at 150° [ref]. Nuclear reactions $^{16}O(d,p_1)^{17}O$, $^{16}O(d,p_0)^{17}O$, $^{12}C(d,p_0)^{13}C$, $^{14}N(d,\alpha_1)^{12}C$ and $^{14}N(d,\alpha_0)^{12}C$ were used to identify O, C and N in the $Mn_4N$ film.

Magnetization measurements were performed using the RSO module of a Superconducting Quantum Interference Device magnetometer (SQUID, Quantum Design)

Time-resolved magneto-optical Kerr effect (TRMOKE) measurements [ref]: a 783-nm Ti:Sapphire laser produces a train of laser pulses of less than 1 ps in duration at a repetition rate of approximately 80 MHz. The laser is separated into a pump beam and probe beam with a polarizing beam splitter. The pump beam is modulated at 9 MHz with an Electro-Optical Modulator, and a delay stage adjusts the length of the pump optical path. Meanwhile, the probe beam is modulated with a mechanical chopper at a frequency of 200 Hz. Both pump and probe beams are co-focused on the sample. The size of the beam spot is controlled with objective lenses of varying magnification. The pump and probe fluences are set at ~0.3 and 0.1 mJ cm$^{-2}$, respectively.

**Results**

Crystalline structure

To relax the 8.3% lattice mismatch between MgO substrate and Mn4N thin films, we use Pd as the buffer layer because the lattice constant of Pd, 0.389 nm, is very close to Mn4N. Samples, MgO (substrate)/Pd (0, 2, 8 nm)/Mn4N (40 nm)/Ta (3 nm), were grown to investigate the effect of the crystalline structure of Mn4N by the buffer layer.

Figure 1 shows the crystalline structure for those samples. The $\theta$–$2\theta$ scan of XRD shows the substrate MgO (002) peak and also Mn4N (001) and (002) peaks for all samples (see Fig. 1(a)), indicating a good out-of-plane texture, though a large lattice mismatch exists for the sample without buffer layer. Other MgO [00 *l*] lines are indicated by an asterisk, *. The inset is the zoom-in view of Mn4N (002) peak, one can see that the peak shifts to left with a thick Pd buffer layer meaning that the out-of-plane lattice constant increases. The full width at hall maximum (FWHM) of Mn4N (002) peak is extracted from the rocking curve, as shown in Table 1. The value decreases from 1.21° to 0.87° after depositing buffer layer suggesting an improvement of the crystalline quality of Mn4N thin films. We also obtain the Mn4N (111) peak with in-plane XRD as shown in Fig.1(b). The intensity of Mn4N (111) peak is significantly enhanced by the Pd buffer layer which is another evidence for the improvement of crystalline quality. Furthermore, the X-ray phi scans from 0° to 360° (Fig. 1(c)) display a fourfold symmetry of the Mn4N (220) and (111) peaks for the Mn4N with 8 nm Pd buffer layer. There is no rotation of (220) peaks and a 45° rotation of (111) peaks relative to the MgO substrate (220) peaks, confirming the epitaxial relationship of MgO(001)[100]||Mn4N(001)[100].

Table 1 summaries the out-of-plane lattice constant (*c*), in-plane lattice constant (*a*) and the ratio of *c*/*a*, respectively. *c* and *a* were extracted from the Mn4N (002) and (111) peaks. As the lattice constant of MgO is larger than Mn4N resulting in a tensile stress, the *c*/*a* ratio of Mn4N without Pd buffer layer is smaller than 1. By increasing the thickness of Pd buffer layer, this ratio increases from 0.987 to 0.992 meaning that the tensile stress is relaxed. Overall, the crystalline structure of Mn4N is improved by the Pd buffer layer.

Stoichiometry

As the reaction of Mn and N atoms can resulting in various phases with different composition affecting the magnetization [J. Appl. Phys. 117, 17B512 (2015)]. It is necessary to check the composition of Mn and N. We employ ion beam analysis including Rutherford backscattering spectrometry (RBS) and nuclear reaction analysis (NRA) which have a high sensitivity to probe light elements, such as N and O [https://doi.org/10.1063/1.3033682]. Fig. 2(a) shows RUMP fitted RBS spectrum of Mn$_4$N film deposited on MgO substrate. RBS spectrum shows Ta from capping layer, Mn peak from Mn$_4$N layer, Mg and O from MgO substrate with N signal overlapping with the substrate signal. The RUMP fitting of the RBS spectrum yields areal density of $2.5 \times 10^{17}$ atoms/cm$^2$ and Mn and N stoichiometry as Mn:N = 4:1. Also, the fitting gives the thickness of Mn4N with 40.5 nm. Fig. 2(b) shows NRA spectrum comprising O peaks from MgO substrate, C from surface hydrocarbons, and N in the Mn₄N film. NRA results confirmed the presence of N in the Mn₄N film which is consistent with RBS results.

Magnetic properties

Magnetic hysteresis loops of 40 nm Mn4N thin films were measured at 300 K under an external magnetic field along the out-of-plane (OP) and in-plane (IP) directions, respectively, as shown in Figs. 3(a) and (b). All samples exhibit a square loop along OP direction but non-hysteresis curve along IP direction suggesting that these samples show PMA. For the Mn4N thin film without buffer layer, the ratio of remanent magnetization ($M_r$) and saturation magnetization ($M_s$), $M_r/M_s$, is 0.66 with 0.41 T coercive field $\mu_0 H_c$. With 8 nm Pd buffer layer, this ratio increases to 0.91 with $\mu_0 H_c = 0.78$ T displaying a significant improvement of PMA.

The parameters of $M_s$, $\mu_0 H_c$ and $K_{eff}$ are summarised in Fig. 3(c), where $K_{eff}$ is the effective magnetic anisotropy constant. $M_s$ for Mn4N thin film without buffer layer is 80.2 kA/m which agrees with the value from literatures [J. Appl. Phys. 117, 17B512 (2015), Adv. Electron. Mater. 2022, 8, 2100772]. However, $M_s$ decreases to 47.6 kA/m by inserting an 8 nm Pd buffer layer. It is known that Mn$_I$ atoms located at the cubic corner are ferromagnetic coupling with each other, while Mn$_{II}$ atoms located at the face center are noncollinear antiferromagnetic coupling with each other. Eventually, the Mn$_I$ atoms anti-parallels to the Mn$_{II}$ atoms resulting in the total magnetization of Mn4N, as shown in Fig. 3(d). It has been reported that a large lattice constant of Mn4N translates to a large atomic separation in the cubic crystal, leading to an increase of Mn-Mn exchange interaction, and thus resulting an increase of the magnetization of Mn$_{II}$ atoms while this effect does not influence Mn$_I$ atoms [Acta Materialia 234 (2022) 118021]. Based on this observation, the out-of-plane lattice constant increases with Pd buffer layer implying an enhancement of magnetization for the Mn$_{II}$ atoms. Thus, total magnetization $M_s$ with buffer layer decreases due to the Mn$_I$ and Mn$_{II}$ atoms anti-parallel with each other. Also, this could be the reason for the enhancement of PMA because a stronger Mn-Mn exchange interaction can lead to a larger $\mu_0 H_c$.

The effective magnetic anisotropy constant $K_{eff} = \frac{1}{2}\mu_0 H_k M_s$ is 4.4, 5.3 and 6.6×10⁴ J m⁻³ for Mn4N with 0 nm, 2nm and 8nm Pd buffer layer, respectively. The anisotropy field $\mu_0 H_k$ is obtained from in-plane saturation magnetic field. Overall, the PMA of Mn4N thin films is enhanced by the Pd buffer layer.

Spin dynamics

To investigate spin dynamics of Mn4N thin films, TR-MOKE measurements were conducted on Mn₄N (30 nm) thin film with 8 nm Pd buffer layer. The schematic of TR-MOKE geometry is illustrated in Fig. 4(a) with the definition of magnetization ($M$), external magnetic field ($\mu_0 H_{ext}$), and corresponding angles, respectively. $\mu_0 H_{ext}$ is fixed at $\theta_H = 75°$ away from the sample surface normal. The measurements were conducted under $\mu_0 H_{ext}$ ranging from 2.25 T to 2.8 T. After an ultrafast laser pulse excites the magnetization, it will precess around the equilibrium direction. Thus, damped oscillating fringes resulting from spin precession can be observed, which allows us to analyze the behavior of the spin dynamics.

A set of raw TR-MOKE signals taken at varying $\mu_0 H_{ext}$ are plotted in Fig. 4(b) as functions of the time delay between pump excitation and probe sensing. The detected TR-MOKE signal is rather weak, presumably due to the low $M_s$ of Mn₄N (47.6 kA/m), considering the Kerr rotation roughly scales with $M_s$ [Nature Photonics, 12, 73–78 (2018)]. Thus, we focus on the analysis of TR-MOKE signals at $\mu_0 H_{ext} \geq 2.25$ T exhibiting sufficient oscillation features. Fig. 4(c) shows precession frequency *vs* $\mu_0 H_{ext}$, extracted from TR-MOKE signals based on $\theta_K = A + Be^{-t/C} + D\sin(2\pi f t + \varphi)e^{-t/\tau}$ [ACS Applied Electronic Materials 3(1), 119-127 (2020), Online ISBN: 978-0-7503-1738-2, *Scientific Reports*, **8**(1), 13395 (2018), *Physical Review Materials*, **6**, 113402 (2022)]. Here, $\theta_K$, *A, B*, and *C* are the TR-MOKE signal, the offset of thermal background, the amplitude of thermal background, and the exponential decaying constant of thermal background, respectively. *D* denotes the amplitude of oscillations and *t* represents time delay between pump and probe beam. *f* and *φ* are precessional frequency and initial phase of precession, respectively. *τ* is relaxation time of precession. A very high frequency of 84 GHz and almost 100 GHz can be observed at 2.25 T and 2.8 T, respectively.

Based on the hysteresis loop measurements, this sample has a PMA. With this constraint condition ($\mu_0 H_k > 0$), the *f* vs $\mu_0 H_{ext}$ relation is fitted to the Kittel formula [*Scientific Reports*, **8**(1), 13395 (2018)]:

$$f = \frac{\gamma}{2\pi} \sqrt{H_1 H_2}, \tag{1}$$

$$H_1 = H_{ext} \cos(\theta - \theta_H) + H_k \cos^2(\theta), \tag{2}$$

$$H_2 = H_{ext} \cos(\theta - \theta_H) + H_k \cos(2\theta), \tag{3}$$

$$2 H_{ext} \sin(\theta_H - \theta) = H_k \sin(2\theta). \tag{4}$$

where *g* is gyromagnetic ratio. With $\theta_H$ fixed at 75°, it shows a reasonable fitting (Fig. 4(c), red curve) and yields an effective anisotropy field of $\mu_0 H_k = 1.74 \pm 0.15$ T. The effective damping $a_{eff} \approx 0.2$ for $\mu_0 H_{ext} \geq 2.25$ T as depicted in Fig. 4(d). Gilbert damping is not extracted due to the limited data points and large uncertainties.

Discussion

Regarding to the large $a_{eff}$, it can be induced by spin pumping effect in which the spin current generated by a precessing magnetization is absorbed by the non-magnetic material [PRL 88, 117601 (2002)]. The strength of the spin pumping effect depends on the spin absorption efficiency of the non-magnetic layer, which is proportional to $Z^4$, where *Z* is the atomic number. In our sample. the Mn4N thin film is sandwiched between Pd and Ta with large Z which could be one of the reasons for the observation of large $a_{eff}$.

We should address that the precession frequency of Mn4N is significant. The frequency obtained in Mn4N thin films is two times higher compared with the traditional CoFeB PMA thin films at same external magnetic field [*Scientific Reports*, **8**(1), 13395 (2018)]. It is also higher than other compensated ferrimagnet at room temperature, such as GdCoFe [Phys. Rev. B 73, 220402(R) (2006)], $Mn_2Ru_xGa$ [PHYSICAL REVIEW B 100, 104438 (2019)]. It is worth to note that the angular momentum in ferrimagnetic Mn4N thin films is not fully compensated, which means the frequency can be enhanced even higher at compensated point. For instance, by doping Ni, zero angular momentum can be achieved in $Mn_{4-x}Ni_xN$ [Nano Lett. 2021, 21, 2580−2587].

Summary

In summary, ferrimagnetic Mn4N thin films with high Curie temperature have been prepared on MgO substrate. The crystalline quality of Mn4N is enhanced significantly by Pd buffer layer to relax the tensile stress from substrate. Ion beam analysis is employed to confirm the Mn:N = 4:1 stoichiometry and thickness. Furthermore, an enhancement of PMA and reduction of magnetization are observed for the Mn4N thin films with Pd buffer layer. The reason for both is an increase of Mn-Mn exchange, and thus resulting an increase of magnetization of the $Mn_{II}$ atoms due to the expansion of the out-of-plane lattice constant. Finally, a fast spin precession frequency, almost 100 GHz, in Mn4N thin films has been demonstrated. Combining high thermal stability, enhanced PMA by buffer layer and fast spin precession, Mn4N thin film is a promising material for spintronic applications.

Figure caption

Figure 1: XRD measurements for Pd (0, 2, 8 nm)/Mn4N (40 nm) thin films. (a) XRD $\theta$–$2\theta$ scan, the inset is the Mn4N (002) peak. (b) In-plane XRD $\theta$–$2\theta$ scan for Mn4N (111) peak. (c) In-plane phi scans from 0° to 360° for Pd (8 nm)/Mn4N (40 nm) sample.

Table 1: The summary of the FWHM, $c$, $a$, and the ratio of $c/a$, respectively.

Figure 2: Rutherford backscattering spectrometry and nuclear reaction analysis for 40 nm Mn4N thin film. The red curve is the RUMP fitting.

Figure 3: (a)-(b) Hysteresis loops for Pd (0, 2, 8 nm)/Mn4N (40 nm) thin films under the magnetic field along out-of-plane and in-plane direction, respectively. (c) The parameters of $M_s$, $\mu_0 H_c$ and $K_{eff}$ for Mn4N thin films with various Pd buffer layer. (d) The schematic of the coupling for $Mn_I$ and $Mn_{II}$ atoms.

Figure 4: (a) Sample stack configuration for TR-MOKE measurement and schematics for physical parameters. $\theta$ and $\theta_H$ denote the equilibrium position of magnetization and the direction of external magnetic field. (b) Sets of TR-MOKE data (symbol) as a function of time delay under varying $\mu_0 H_{ext}$ along $\theta_H = 75°$. Solid curves are the best fits. (c)-(d) Precession frequency $f$ and $\alpha_{eff}$ as a function of $\mu_0 H_{ext}$, respectively. The red curve is the Kittel formula fitting.